\documentclass[twocolumn,secnumarabic,amssymb, nobibnotes, aps, pra,showpacs,preprintnumbers]{revtex4-1}
\usepackage{amssymb}
\usepackage{array}
\usepackage{graphicx}
\usepackage{amsmath}
\usepackage{amsthm}
\usepackage{amsopn}
\def\uno{\mbox{1 \kern-.59em {\rm l}}}

\def\beq{\begin{equation}}
\def\eeq{\end{equation}}
\def\bea{\begin{eqnarray}}
\def\eea{\end{eqnarray}}

\begin{document}

\markboth{Authors' Names}
{Instructions for Typing Manuscripts (Paper's Title)}

\title{A Dynamical System Analysis of $f(R,T)$ Gravity
}

\author{Behrouz Mirza}
\email{b.mirza@cc.iut.ac.ir}
\author{Fatemeh Oboudiat}
\email{f.oboudiat@ph.iut.ac.ir}
\affiliation{Department of Physics,
Isfahan University of Technology, Isfahan 84156-83111, Iran}

\begin{abstract}
We investigate equations of motion and future singularities of $f(R,T)$ gravity where $R$ is the Ricci scalar and $T$ is the trace of stress-energy tensor. Future singularities for two kinds of equation of state (barotropic perfect fluid and generalized form of equation of state) are studied. While no future singularity is found for the first case,  some kind of singularity is found to be possible  for the second. We also investigate $f(R,T)$ gravity by the method of dynamical systems and obtain some fixed points. Finally, the effect of the Noether symmetry on $f(R,T)$ is studied and the consistent form of $f(R,T)$ function  is found using the symmetry and the conserved charge.
\end{abstract}

\maketitle

\section{Introduction}
Recent cosmological observations show that our universe has an accelerating expansion \cite{astro}. Two groups of solutions are available that can be invoked for explaining the phenomenon. The first one is based on the belief that some exotic matter exist within the framework of General Relativity (GR) known as dark energy  that has the parameter $\omega <0$ in its equation of motion.  Such a matter raises some fundamental questions such as the existence of negative entropy,  future singularities, and the violation of some energy conditions.
\\On the other hand, some authors have generalized GR to some new theories of gravity \cite{modified}. Finite time future singularities of these theories was studied in \cite{fs}. One of these is $f(R)$ gravity in which the standard Einstein-Hilbert action is replaced with an arbitrary function of Ricci scalar \cite{rev f(R)}. In addition to its capability to describe the expansion of the universe without introducing any dark energy \cite{f1},  this generalized theory of gravity  has other advantages. For example, it can explain the dynamics of galaxies without recourse to the concept of dark matter \cite{f2} and unifies inflation with dark energy \cite{f3,f4,f5,f6,f7}.
\\A further generalization of $f(R)$ is $f(R,T)$ gravity where $T$ is the trace of stress-energy tensor \cite{Harko}. As a consequence of using stress-energy tensor as a source, the motion of the particles does not take place along a geodesic path because there is an extra force perpendicular to the four-velocity unless we add the constraint of conservation of stress-energy tensor (unlike GR and $f(R)$ theories, the  continuity equation is independent of equations of motion in this case). It is shown in \cite{perturbation} that due to the conservation of stress-energy tensor, $T$ sector of $f(R,T)$ cannot be chosen arbitrarily but it has a special form. The thermodynamics of this model is studied in \cite{sharif}, and the possibility of wormhole geometry is examined  in \cite{azizi}. Also energy conditions \cite{frt energy}, cosmological solutions \cite{cosmo}, scalar perturbations \cite{perturbation} are investigated. In \cite{farhoodi} solar system consequences of the model is argued. Further generalization of this theory to $f(R,T,R_{\mu\nu}T^{\mu\nu})$ is proposed in \cite{rmunu}.
\\The method of autonomous dynamical systems is a useful tool for investigating the modified theories of gravity such as $f(R)$ and some other theories \cite{rev fixed,fixfr,fix,biswas}. In \cite{fix frt} the method is investigated for $f(R,T)$ theory assuming conservation of energy. In this paper we study the method by no limiting condition on energy.
\\The concept of symmetry has always been an attractive subject in physics. Noether symmetry attracts more attention because it helps to find constants of motion (like energy and momentum) from continuous symmetries of the system. Some efforts has been done to look for such conserved quantities in cosmological models \cite{Noether}. Some  authors have considered the effect of Noether symmetry in extended theories of gravity such as $f(R)$ \cite{fr1} and $f(T)$ theories \cite{ft}. We investigate the effect of Noether symmetry on $f(R,T)$ theory to see if it is possible to make a consistent form of $f(R,T)$ by Noether symmetry.
\\This article is organized as follows. In section 2, we briefly explain the action of the $f(R,T)$ model and obtain the gravitational field equations. In section 3, we consider singularities for dark energy. The aim of section 4 is to consider the method of dynamical systems in $f(R,T)$ theory. In section 5 the Noether symmetry is studied. Section 6 concludes with a  summary and discussion.

\section{Equations of motion}
  General form of the action for $f(R,T)$ model is:
\bea
S=\int d^{4}x \sqrt{-g}\left\{\frac{1}{2k}f(R,T)+ \mathcal{L}_{m}\right\}
\eea
where $k=8\pi G$, $G$ is Newtonian constant, $R$ is Ricci scalar, $T$ is the trace of stress-energy tensor and $\mathcal{L}_{m}$ is matter lagrangian density. By varying the action with respect to the metric we can find equations of motion:
\bea
&&f_{R}R_{\mu\nu}-\frac{1}{2}f(R,T)g_{\mu\nu}+(g_{\mu\nu}\nabla_{\alpha}\nabla^{\alpha}-\nabla_{\mu}\nabla_{\nu})f_{R}\nonumber\\
&&=k T_{\mu\nu}-f_{T}(T_{\mu\nu}+\Theta_{\mu\nu})\label{motion}
\eea
where:
\bea
&\Theta_{\mu\nu}=g^{\alpha\beta}\frac{\delta T_{\alpha\beta}}{\delta g_{\mu\nu}}=-2T_{\mu\nu}+g_{\mu\nu}\mathcal{L}_{m}-2g^{\alpha\beta}\frac{\partial^{2}\mathcal{L}_{_{m}}}{\partial g^{\mu\nu}\partial g^{\alpha\beta}}&\nonumber\\
& f_{R}=\frac{\partial f(R,T)}{\partial R}\quad f_{T}=\frac{\partial f(R,T)}{\partial T}&
\eea
The following equations can also be obtained easily
\bea
&\nabla^{\mu}G_{\mu\nu}=0 &\label{1}\\
&\nabla^{\mu}(\nabla_{\mu}\nabla_{\nu}-g_{\mu\nu}\nabla_{\alpha}\nabla^{\alpha})f(R,T)=R_{\mu\nu}\nabla^{\mu}f(R,T)&\label{2}
\eea
where $G_{\mu\nu}$ is the Einstein tensor.
By contracting (\ref{motion}) into $\nabla^{\mu}$ and using (\ref{1}) and (\ref{2}) we have:
\bea
&&(T_{\mu\nu}+\Theta_{\mu\nu})\nabla^{\mu}f_{T}+f_{T}\nabla^{\mu}\Theta_{\mu\nu}-\frac{1}{2}g_{\mu\nu}(\nabla^{\mu}T)f_{T}\nonumber\\
&&=\left(k-f_{T}\right)\nabla^{\mu}T_{\mu\nu}=j_{\nu}\label{con22}
\eea
Replacing $f(R,T)$ by  $f(R)$ in equation (\ref{con22}) leads to the  conservation of energy-momentum tensor, $\nabla^{\mu}T_{\mu\nu}=0$. This means that symmetries of  $f(R)$ theory implies conservation of the energy momentum tensor. However, in $f(R,T)$ theory the energy momentum tensor is not generally  conserved. In this and the next section we study $f(R,T)$ by assuming the conservation of energy and in the following sections we investigate more general cases.

\noindent Assuming $\nabla^{\mu}T_{\mu\nu}=0$ in equation (\ref{con22}) we obtain the following equation,
\bea
(T_{\mu\nu}+\Theta_{\mu\nu})\nabla^{\mu}f_{T}+f_{T}\nabla^{\mu}\Theta_{\mu\nu}-\frac{1}{2}g_{\mu\nu}(\nabla^{\mu}T)f_{T}=0. \label{con222}
\eea
 Now, we concentrate on the case $f(R,T)=R+g(T)$. Assuming the matter content of the universe as a perfect fluid $\left(T_{\mu\nu}=(\rho  +p)u_{\mu}u_{\nu}-pg_{\mu\nu}\right)$  and having a FRW universe, i.e.:
\bea
ds^{2}=dt^{2}-a^{2}(t)\left(\frac{dr^{2}}{1-\kappa r^{2}}+d\Omega^{2}\right)
\eea
the Friedman equations become:
\bea
&3H^{2}+\frac{\kappa}{a^{2}}=\left[k+g'(T)\right]\rho +g'(T)p+\frac{1}{2}g(T)&\label{fr1}\\
&-2\dot{H}-3H^{2}-\frac{\kappa}{a^{2}}=k p-\frac{1}{2}g(T)& \label{fr2}
\eea
 The continuity equation is independent from the Friedman equations:
\bea
\dot{\rho}+3H(\rho +p)=0. \label{con}
\eea
Adding the equation of state $p=p(\rho)$ to the three equations (\ref{fr1}),(\ref{fr2}) and (\ref{con}), we have four independent equations which can be used to obtain the time dependence of four parameters of $\rho,p,a,$ and $g(T)$. To do this for flat universe ($\kappa =0$) we add  the Friedman equations (\ref{fr1}) and (\ref{fr2}):
\bea
-2\dot{H}=\left[k+g'(T)\right](\rho+p) \label{fr3}
\eea
Eliminating the term $\rho +p$  from (\ref{con}) and (\ref{fr3}), we have:
\bea
6H\dot{H}=\left[k+g'(T)\right]\dot{\rho}\label{hh}
\eea
By differentiating (\ref{fr1}) and substituting (\ref{hh}), we have:
\bea
\dot{g}'(T)(\rho +p)+\dot{p}g'(T)+\frac{1}{2}\dot{g}(T)=0 \label{diffg}
\eea
By solving this equation for a perfect barotropic fluid ($p=\omega\rho$),  we have for $\omega\neq\pm\frac{1}{3},-1$:
\bea
g(T)=g_{0}T^{\theta},\quad\theta=\frac{1+3\omega}{2(1+\omega)}\label{abc}
\eea
Also we found equation (\ref{abc}) for special case $f(R,T)=R+g(T)$ but it is possible to find the same relation for general case $f(R,T)=h(R)+g(T)$ from equation (\ref{con22}). The Friedman equations become:
\bea
3H^{2}&=&k\rho+g_{0}(1-3\omega)^{\alpha-1}\rho^{\alpha}=k(\rho+\rho_{DE})\\
2\dot{H}+3H^{2}&=&-k p+\frac{1}{2}g_{0}(1-3\omega)^{\alpha}\rho^{\alpha}=-k(p+p_{DE})\quad\quad
\eea
We can interpret the above equations as the sum of matter fluid and DE in the framework of GR where density and pressure of DE are given by:
\bea
\rho_{DE}&=&g_{0}(1-3\omega)^{\alpha-1}\rho^{\alpha}\\
p_{DE}&=&-\frac{1}{2}g_{0}(1-3\omega)^{\alpha}\rho^{\alpha}=-\frac{1}{2}(1-3\omega)\rho_{DE}
\eea
Both fluids are, therefore, perfect with the equation of state parameter $\omega$ and $\omega_{DE}=-\frac{1}{2}(1-3\omega)$ \cite{Chak}. To investigate the future singularities of this model, we first solve the continuity equation (\ref{con}) as follows:
\bea
\rho=\rho_{0}a^{-3(1+\omega)}\label{ro}
\eea
By substituting this equation in Friedman equation (\ref{fr1}), we have:
\bea
\pm(t-t_{0})=\int\frac{a^{\frac{1+3\omega}{2}}da}{\sqrt{d_{1}+d_{2}a^{\frac{3(1-\omega)}{2}}}}
\eea
where, $d_{1}=\frac{k\rho_{0}}{3}$ and $d_{2}=\frac{g_{0}\rho_{0}^{\alpha}(1-3\omega)^{\alpha-1}}{3}$. By substituting different allowed values of $\omega$ ($\omega\neq\pm\frac{1}{3},-1$) in the above equation, we find no future singularities in this model. Below, we will consider the Friedman equations and future singularities for a more general equation of state.
\\We can interpret the Friedman equations (\ref{fr1}) and (\ref{fr2}) in a different way. If we define $\rho_{de}$ and $p_{de}$ and $\tilde{k}$ as:
\bea
\rho_{de}=-p_{de}&\equiv &\frac{pg'(T)+\frac{1}{2}g(T)}{\widetilde{k}}\\
\widetilde{k}&\equiv & k+g'(T)\equiv 8\pi \widetilde{G}
\eea
then, the Friedman equations become:
\bea
3H^{2}&=&\widetilde{k}(\rho +\rho_{de})\label{f1}\\
-3H^{2}-2\dot{H}&=&\widetilde{k}(p+p_{de})\label{f2}
\eea
From above discussions it is obvious that there are some special features that we can distinguish this model from any typical dark energy model. One of the important features is that $f(R,T)$ does not preserve conservation of energy momentum tensor and continuity equation but it is possible to make the model to preserve it. The coupling constant can be constant (equations (\ref{fr1}) and (\ref{fr2})) or have running with energy (equations (\ref{f1}) and (\ref{f2})) like field theories QED, QCD, ... furthermore $\rho_{de}$ and $p_{de}$ have the behavior of dark energy so we can explain expanding of the universe without introducing any exotic matter like dark energy.
\section{Generalized equation of state}

In this section, we  consider a more general equation of state:
\bea
p=-\rho-f(\rho) \label{state}
\eea
This kind of equation of state leads to five types of singularities in $f(R)$ theory \cite{sin1,sin2}:
\begin{itemize}
  \item Type I ("Big Rip"): $t\rightarrow t_{s}, a\rightarrow \infty, \rho\rightarrow\infty,$ and $|p|\rightarrow\infty$
  \item Type II ("Sudden"): $t\rightarrow t_{s}, a\rightarrow a_{s}, \rho\rightarrow\rho_{s},$ and $|p|\rightarrow\infty$
  \item Type III: $t\rightarrow t_{s}, a\rightarrow a_{s}, \rho\rightarrow\infty,$ and $|p|\rightarrow\infty$
  \item Type IV: $t\rightarrow t_{s}, a\rightarrow \infty, \rho\rightarrow0,$ and $|p|\rightarrow0,$ but higher derivatives of $H$ diverges.
  \item Type V: In this type of singularity, $\omega =\frac{p}{\rho}$ diverges and it is possible that none of the other parameters has a  singularity.
\end{itemize}
In the following, we will try to see  if similar types of singularity exist in the $f(R,T)$ model. Assuming (\ref{state}) and  from the continuity equation (\ref{con}), we have:

\bea
\dot{\rho}=3Hf(\rho) \label{con2}
\eea
and from (\ref{diffg}), we have:
\bea
g'(T)=g'_{0}\sqrt{f(\rho)}a^{3} \label{gf}
\eea
Eliminating $g'(T)$ between (\ref{fr3}) and (\ref{gf}) we have:
\bea
2\dot{H}=\left(k+g'_{0}\sqrt{f(\rho)}a^{3}\right)f(\rho) \label{af}
\eea
Having $a(t)$, we can now get the behavior of $\rho(t),p(t),g(t)$ and $f(\rho(t))$ from (\ref{state}),(\ref{con2}),(\ref{gf}) and (\ref{af}). If we assume $H(t)$ to have a singular form as:
\bea
H(t)=h(t_{s}-t)^{-m}
\eea
where $t_{s}$ is the time of future singularity then for $m = 1$, $a(t)$ becomes:
\bea
a(t)=a_{0}(t_{s}-t)^{-h}.\label{scale factor}
\eea
The behavior of other functions near $t_{s}$ will be  as follows:
\bea
\rho ,p&\propto&(t_{s}-t)^{\alpha}\\g&\propto&(t_{s}-t)^{\beta}\\g'&\propto&(t_{s}-t)^{\gamma}
\eea
The values of $\alpha$, $\beta$ and $\gamma$ for different values of $h$ are shown in Table \ref{tab1}.
\begin{table}
\begin{center}
\caption{\bf  The values of exponents of $\rho,p,g,$ and $g'$ for different values of $h$}
\vspace{5mm}
\begin{tabular}{|c|c|c|c|}
\hline
{\bf h}&  {\bf$\alpha$}& {\bf $\beta$} & {\bf $\gamma$} \\[0.3ex]
\hline

3&$\frac{14}{3}$&$-\frac{20}{3}$&$-\frac{5}{3}$\\
&&&\\
2&$\frac{8}{3}$&$-\frac{14}{3}$&$-\frac{11}{3}$\\
&&&\\
1&$\frac{2}{3}$&$-\frac{8}{3}$&$-\frac{17}{3}$\\
&&&\\
-1&-6&0&8\\
&&&\\
-2&-12&0&17\\
&&&\\
-3&-18&0&26\\

\hline
\end{tabular}
\label{tab1}
\end{center}
\end{table}

\noindent For $h>0$, $g$ and $g'$ have singular behaviors near $t_{s}$ and $\beta$ and $\gamma$ are their exponents but $\rho$ and $p$ are finite and their exponents are $\alpha$. For $h<0$, $\rho$ and $p$ are singular with the exponent $\alpha$ while  $g$ and $g'$ are finite at $t_{s}$. The following relations holds between the exponents:
\bea
h>0&:&\beta_{h}=\frac{\alpha_{h}}{2}-3h,\quad\alpha_{h}=-\beta_{h-1},\quad\gamma_{h}=\beta_{h}-1\quad\quad\\
h<0&:&\beta_{h}=0,\quad\alpha_{h}=6h,\quad\gamma_{h}=3h-2
\eea
For $m\neq1$, $a(t)$ has the following different form:
\bea
a(t)=a_{0}exp\left[ \frac{h(t_{s}-t)^{1-m}}{m-1}\right]
\eea
Depending on $m$, we have different results near $t_{s}$:
\begin{enumerate}
  \item $m<-1$: It is obvious from the form of $a,H,\dot{H}$ that none of them are singular near $t_{s}$ and it is also clear  from (\ref{con2}),(\ref{gf}) and (\ref{af}) that $f(\rho), g'(T), \dot{\rho}$ are finite, too. In this case, we have no singularity in $t_{s}$ except for the higher derivatives of $H$.

  \item $-1<m<0$: In this case, at $t_{s}$, $a$ and $H$ are finite and just $\dot{H}$ is singular, it is obvious  from (\ref{con2}),(\ref{gf}) and (\ref{af}) $f(\rho),g'(T),\dot{\rho}$ are singular. We can determine the behavior of $\rho,p$ and $g$ by numerical solution of equations (\ref{state}),(\ref{con2}),(\ref{gf}) and (\ref{af}).

  \item $0<m<1$: In this case, $H$ and $\dot{H}$ are singular but $a$ is finite. From (\ref{gf}) and (\ref{af}) $f$ and $g'$ must be singular. Again, we see that all  the parameters have singular behavior near $t_{s}$ but $p$ is negative in this case.

  \item $m>1$: In this case, all the  three  $a,H,\dot{H}$ are singular. In this case, just $g$ and $g'$ are singular near $t_{s}$. $\rho$ and $p$ tend to zero from below.

\end{enumerate}

\begin{table*}
\begin{center}
\caption{\bf Singularity of cosmological parameters for the Hubble parameter $H(t)=h(t_{s}-t)^{-m}$ }
\vspace{5mm}
\begin{tabular}{|c|c|c|}
\hline
{\bf values of  $m$}& {\bf values of scale factor} &{\bf values of other parameters} \\\hline &&\\
$m<-1$& $a$, $H$ and $\dot{H}$ are finite. & $\rho$, $p$, $g$ and $g'$ are finite.\\&&\\ \hline &&\\
$-1<m<0$&$\dot{H}$ is singular. & $\rho$, $p$, $g$ and $g'$ are singular.\\ &&\\\hline &&\\
$0<m<1$&$H$ and $\dot{H}$ are singular. & $\rho$, $p$, $g$ and $g'$ are singular.\\&&\\ \hline &&\\
$m=1$ \begin{tabular}{c}

       $h>0$ \\
       $h<0$ \\

     \end{tabular}
&\begin{tabular}{c}

                   $a$, $H$ and $\dot{H}$ are singular. \\
                   $H$ and $\dot{H}$ are singular. \\

                 \end{tabular}
     &\begin{tabular}{c}

       $g$ and $g'$ are singular, $\rho$ and $p$ are finite. \\
       $g$ and $g'$ are finite, $\rho$ and $p$ are singular. \\

      \end{tabular}
     \\&&\\ \hline &&\\
$m>1$& $a$, $H$ and $\dot{H}$ are singular. & $g$ and $g'$ are singular, $\rho$ and $p$ are finite.\\&&\\\hline

\end{tabular}
\label{tab2}
\end{center}
\end{table*}

We have summarized the above discussion about singularity in Table \ref{tab2}. So we have following types of singularity in $f(R,T)$ model:
\begin{itemize}
  \item Type $\widetilde{I}$: $t\rightarrow t_{s}; H\rightarrow \infty, g\rightarrow \infty, \rho\rightarrow \infty, p\rightarrow \infty, g'\rightarrow \infty$; (for $0<m<1$).
  \item Type $\widetilde{II}$: $t\rightarrow t_{s}; g\rightarrow \infty, \rho\rightarrow \infty, p\rightarrow \infty, g'\rightarrow \infty$; (for $-1<m<0$).
  \item Type $\widetilde{III}$: $t\rightarrow t_{s}; H\rightarrow \infty, g\rightarrow \infty, g'\rightarrow \infty;\rho,p$ finite (for $m=1,h>0$ and $m>1$).
  \item Type $\widetilde{IV}$: $t\rightarrow t_{s}; H\rightarrow \infty, \rho\rightarrow \infty, p\rightarrow \infty;g,g'$ finite (for $m=1,h<0$).
\end{itemize}
It should be noted that the behaviors of $g$ and $g'$ exhibit the new characteristic of these types of singularities.
\section{$f(R,T)$ and fixed points}
Calculations of the previous section was a little restrictive. A special form of scale factor (\ref{scale factor}) was chosen in order to study future singularities of the theory and  conservation of energy was imposed by hand. In this section we will consider a more general scenario. No special form of scale factor is chosen, the continuity equations are  generalized to the following form:
\bea
& &\dot{\rho}_{m}+3H(\rho_{m}+p_{m})=Q\nonumber\\
& &\dot{\rho}_{r}+3H(\rho_{r}+p_{r})=Q'\label{contin}\\
& &\dot{\rho}_{T}+3H(\rho_{T}+p_{T})=-Q-Q'\nonumber
\eea
where, the indices $m$ and $r$ means matter and radiation,  $\rho_{T}=\frac{1}{k}\left[g'(T)\rho_{m}+\frac{1}{2}g(T)\right]$ and $p_{T}$ is equal to $-\frac{g(T)}{2k}$. It should be noted that  $T$ is the trace of energy momentum tensor of both matter and radiation; since energy momentum tensor of radiation is traceless ($p_r=\frac{1}{3}\rho_r$) $T$ equals to the trace of energy momentum tensor of matter only i.e. $T=T_m$. Equation (\ref{contin}) leads the exchange of energy between matter, radiation and dark energy. Dimension of $Q$ and $Q'$ is like $\dot{\rho}$ or $H\rho$. We can put $H\rho_{m}$, $H\rho_{r}$, and $H\rho_{T}$ or a combination of them instead of $Q$ and $Q'$. We choose a simple case $Q=\alpha H\rho_{m}$ and $Q'=\alpha H\rho_{r}$. So the continuity equations can be written as follows:
\bea
& &\dot{\rho}_{m}+3H(\rho_{m}+p_{m})=\alpha H\rho_{m}\nonumber\\
& &\dot{\rho}_{r}+3H(\rho_{r}+p_{r})=\alpha H\rho_{r}\label{contin1}\\
& &\dot{\rho}_{T}+3H(\rho_{T}+p_{T})=-\alpha (H\rho_{m}+ H\rho_{r})\nonumber
\eea
where $\alpha$ is a constant. The Friedman  equations containing radiation and dust matter ($\omega_m=0$) are as follows:
\bea
&&3H^{2}=\left[k+g'(T)\right]\rho_{m} +\frac{1}{2}g(T)+k\rho_{r}\label{ff1}\\
&&-2\dot{H}-3H^{2}=-\frac{1}{2}g(T)+\frac{k}{3}\rho_{r} \label{ff2}
\eea
where indices $r$ and $m$ stand for radiation and matter. Due to complicated form of the above equations we use the method of autonomous dynamical system to study the problem. By introducing dimensionless parameters:
\bea
&\Omega_{m}=\frac{k\rho_{m}}{3H^{2}} &\nonumber\\
&\Omega_{r}=\frac{k\rho_{r}}{3H^{2}} &\label{dynamical}\\
&X=\frac{g}{6H^{2}} &\nonumber
\eea
and using modified Friedman equations (\ref{ff1}) and (\ref{ff2}) and continuity equations (\ref{contin1}) the autonomous system becomes:
\bea
\frac{d\Omega_{m}}{dN}&=&\Omega_{m}\left(\alpha+\Omega_{r}-3X\right)\nonumber\\
\frac{d\Omega_{r}}{dN}&=&\Omega_{r}\left(\alpha-1+\Omega_{r}-3X\right)\label{aoto}\\
\frac{dX}{dN}&=&\frac{\alpha -3}{2}(1-\Omega_{m}-\Omega_{r}-X)+3X(1+\frac{\Omega_{r}}{3}-X)\nonumber
\eea
where $N=\ln a$. The fixed points and the related physical parameters for the system (\ref{aoto}) are represented in table \ref{tab3}.  Deceleration parameter related in the table is defined as bellow:
\bea
q=-\frac{\ddot{a}a}{\dot{a}^{2}}=-1-\frac{\dot{H}}{H^{2}}=\frac{3}{2}\left(\frac{\Omega_{r}}{3}-X\right)+\frac{1}{2}
\eea
Looking for an accelerating universe we should find minus values for nowadays phase of the universe. The theory has four fixed points. To study the stability of the fixed points one has to study the eigenvalues of the first order perturbation matrix near the critical points which are presented in Table \ref{tab3} by $\lambda_{i}$. The stable and unstable regimes are presented too.

A good cosmological model should contain at least a part of the standard cosmological model which is summarized as bellow \cite{rev fixed}:\\
inflation $\rightarrow$ radiation $\rightarrow$ matter $\rightarrow$ accelerating expansion.\\ Hence any matter and radiation fixed point in the model should be a saddle point and any accelerated phase should be an attractor. \\
The fixed point $P_1$ is hyperbolic (a hyperbolic fixed point is the one whose Jacobian matrix has no zero eigenvalue.) except at the values $\alpha =\pm 3, 4$. It represents dark energy dominated era ($\Omega_m=\Omega_r=0$) and accelerated phase ($q<0$) of the universe and it is stable at the values $-3<\alpha<3$.\\
$P_2$ is dark energy dominated and hyperbolic except at the values $1,\frac{5}{3},-3$ and represents accelerated phase of the universe at the domain $\alpha <1$. It is a stable fixed point in the regime $\alpha <-3$ and unstable for $\alpha > \frac{5}{3}$.\\
$P_3$ is a combination of radiation and dark energy. It can describe accelerated phase of the universe for $\alpha > 2$. There is no stable regime for $P_3$ while it is unstable at $\alpha <\frac{5}{3}$.\\
The only point that have dark energy and matter contribution is $P_4$. It represents accelerated phase of the universe for $\alpha >1$ and it is stable for $\alpha > 3$. The values of $\Omega_m$ and deceleration parameter are both proportional to $1-\alpha$. This means near $P_4$ we have acceleration phase or $\rho_m<0 $ that is violation of weak energy condition. Evolution of the universe depends strongly on the values of the coupling parameter $\alpha$. Stable and unstable fixed points for different regimes of $\alpha$ are summarized in table \ref{tab4}. In each regime the universe starts its evolution from unstable fixed point (as Big Bang) and ends in the stable one. Fig. \ref{fig6} represents trajectories of the phase space for different regimes of table \ref{tab4}. As we see from table \ref{tab4} all stable fixed points are in an acceleration phase in the specified values of $\alpha$ and this is one of the successions of the theory. Matter and radiation dominated fixed points should be  saddle points. The only regime that matter and radiation fixed points are saddle is $\frac{5}{3}<\alpha<3$. In the regime $\frac{5}{3}<\alpha<\frac{10}{3}$, $\Omega_r$ becomes negative and weak energy condition is violated so the preferred regime is $\frac{10}{3}<\alpha<3$. For these values of $\alpha$, the saddle point $P_3$ is radiation dominated and there are some trajectories in Fig. \ref{fig6}e that evolves directly from $P_4$ to the final fixed point $P_1$. This means transition of matter era to the acceleration phase.

\begin{figure*}
\begin{center}
 \begin{tabular}{cc}
  \includegraphics[scale=0.8]{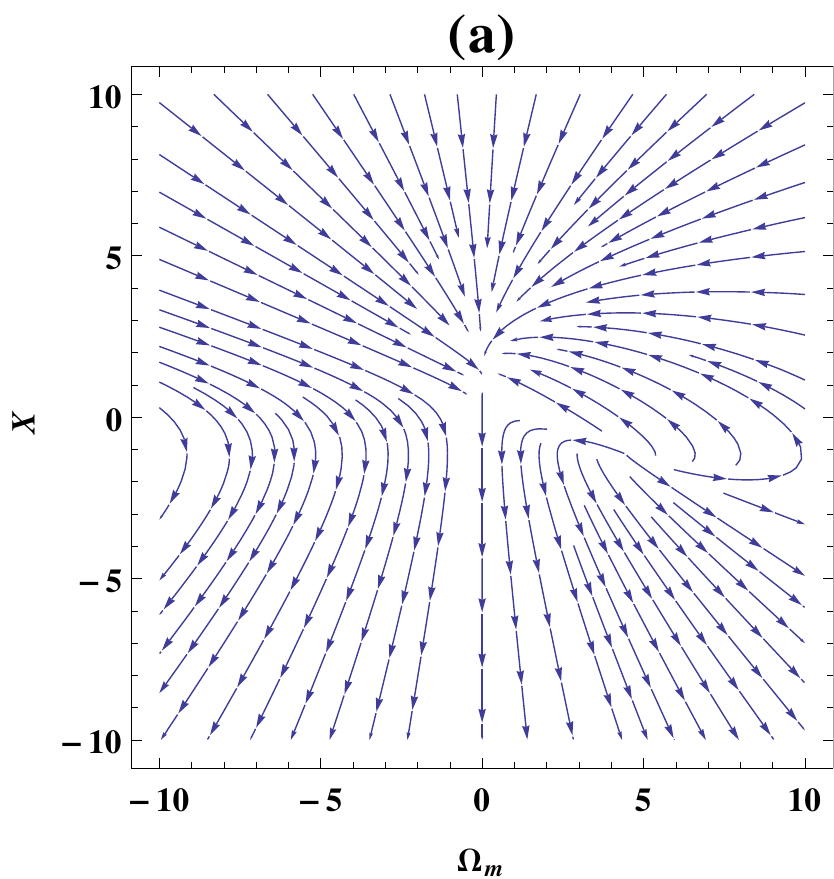}&\includegraphics[scale=0.8]{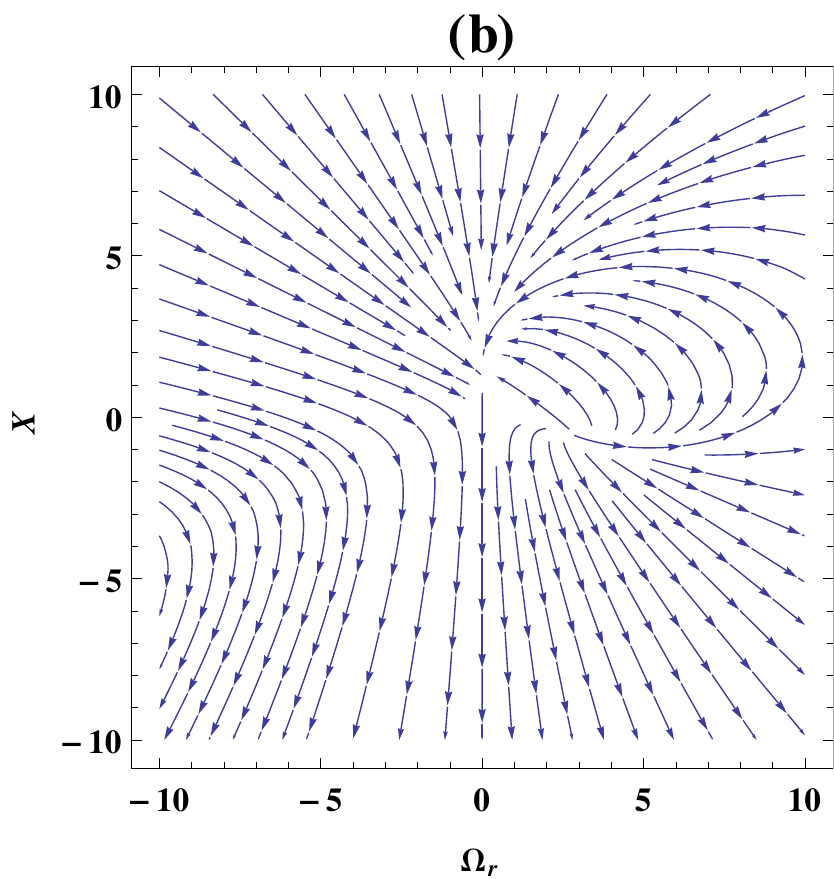}\\
  \includegraphics[scale=0.8]{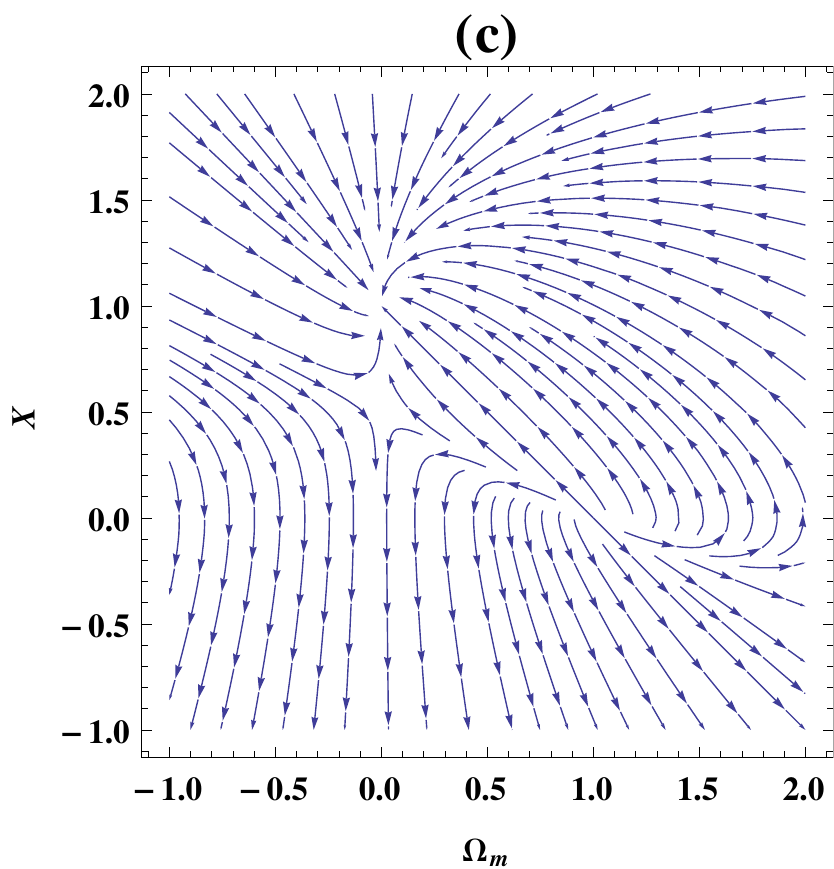}&\includegraphics[scale=0.8]{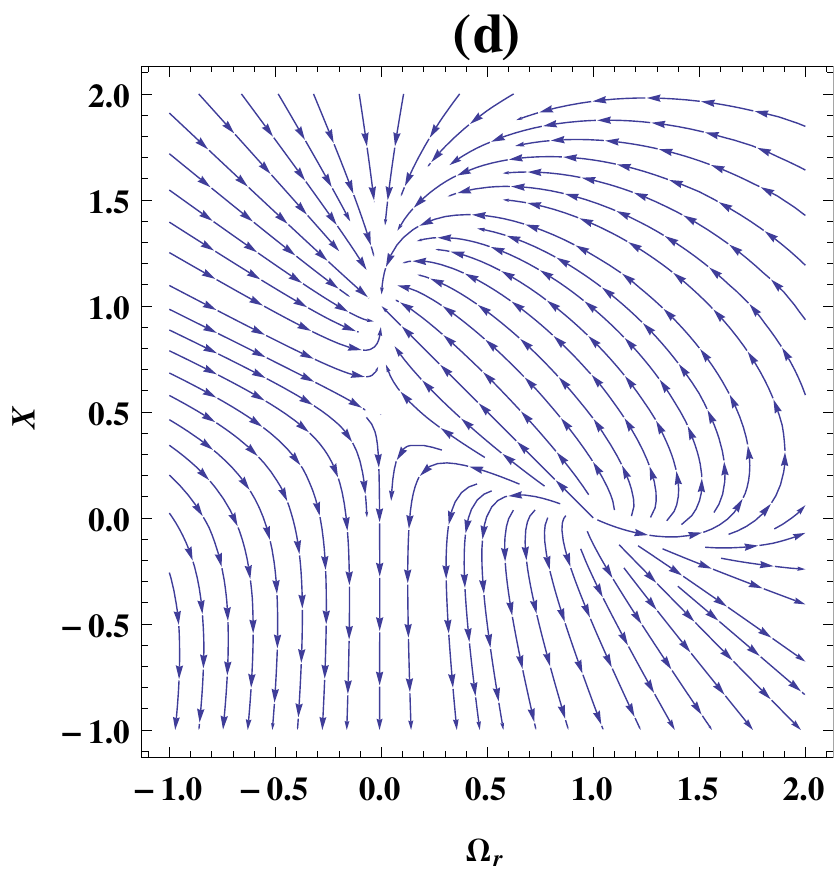}\\
  \includegraphics[scale=0.8]{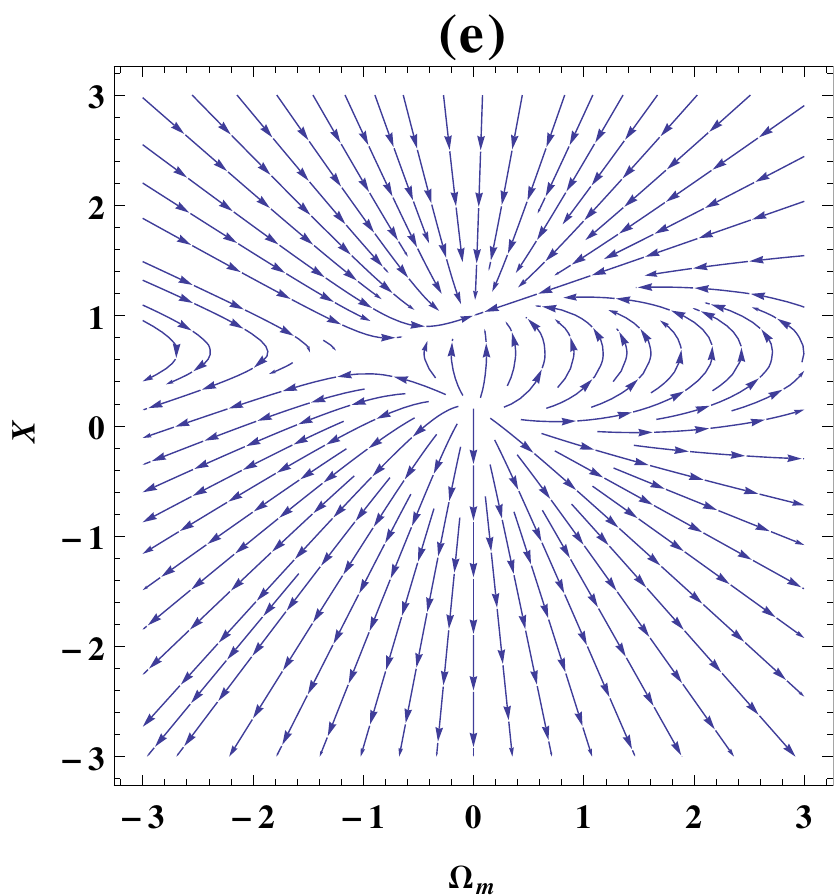}&\includegraphics[scale=0.8]{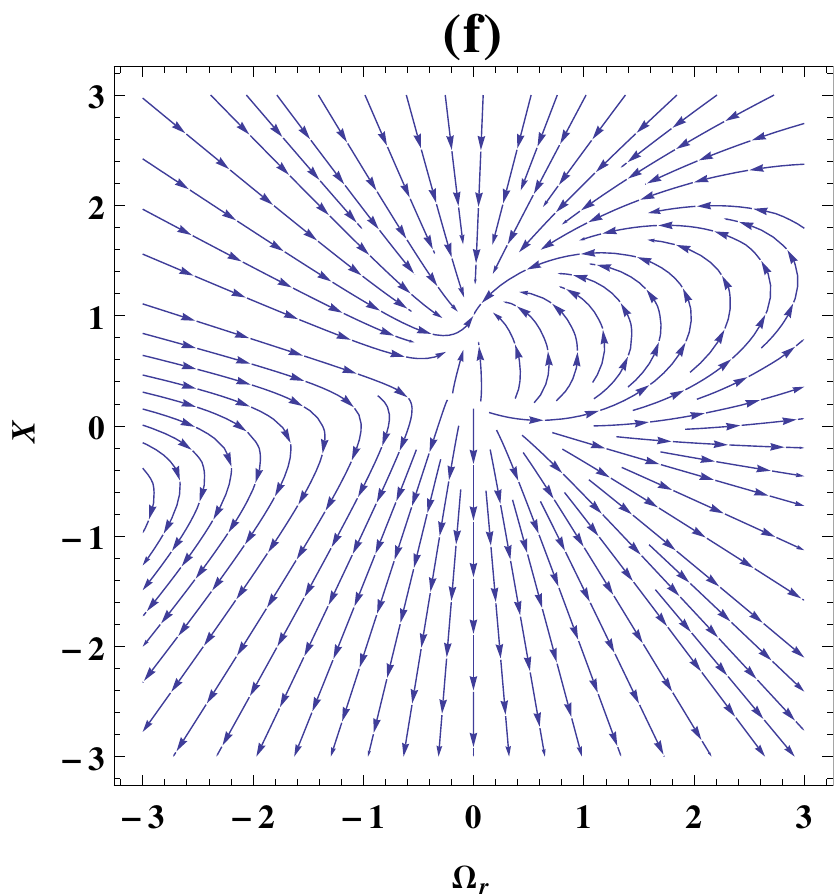}
\end{tabular}
\end{center}
\end{figure*}
\begin{figure*}
\begin{center}
\begin{tabular}{cc}
  \includegraphics[scale=0.8]{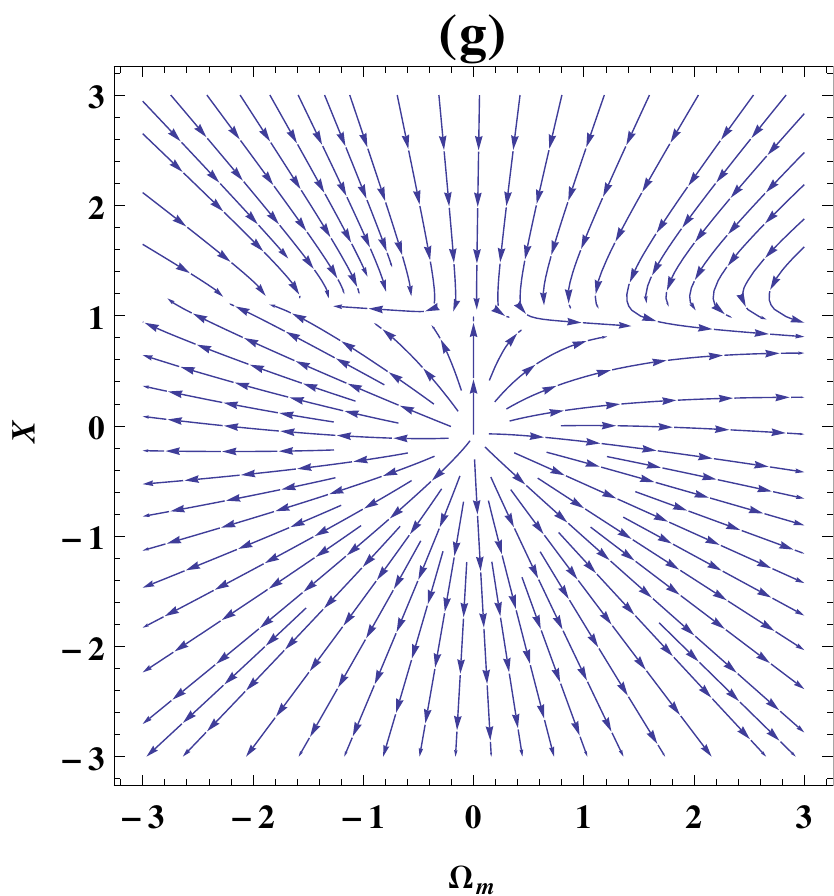}&\includegraphics[scale=0.8]{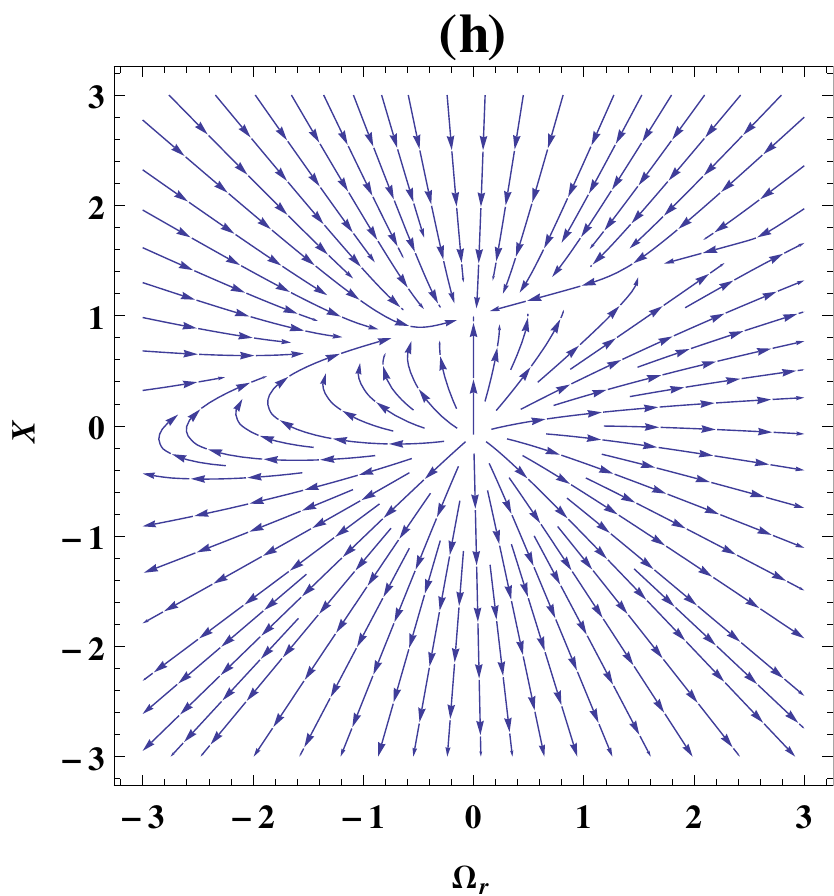}
\end{tabular}
\caption{ The behavior of critical points in phase planes \textbf{(a)} ($\Omega_m,X,\Omega_r=0$) for $\alpha=-3.5$ \textbf{(b)} ($\Omega_r,X,\Omega_m=0$) for $\alpha=-3.5$ \textbf{(c)} ($\Omega_m,X,\Omega_r=0$) for $\alpha=.001$ \textbf{(d)} ($\Omega_r,X,\Omega_m=0$) for $\alpha=.001$ \textbf{(e)} ($\Omega_m,X,\Omega_r=0$) for $\alpha=2$ \textbf{(f)} ($\Omega_r,X,\Omega_m=0$) for $\alpha=2$ \textbf{(g)} ($\Omega_m,X,\Omega_r=0$) for $\alpha=3.5$ \textbf{(h)} ($\Omega_r,X,\Omega_m=0$) for $\alpha=3.5$. Coordinate space of $P_1$ and $P_2$ are at $(0,1)$ and $(0,\frac{3-\alpha}{2})$ in both of the phase planes ($\Omega_r,X,\Omega_m=0$) and ($\Omega_m,X,\Omega_r=0$). $P_3$ is visible in the plain ($\Omega_r,X,\Omega_m=0$) at $(\frac{3\alpha(3-\alpha)}{2(10-3\alpha)}-\alpha+1,\frac{3\alpha(3-\alpha)}{2(10-3\alpha)})$ and $P_4$ is at $(1-\alpha, \frac{\alpha}{3})$ in the plain ($\Omega_m,X,\Omega_r=0$). The transition from the point $P_4(-1,\frac{2}{3})$ to $P_1(0,1)$ in plot \textbf{(e)} represents the transition from matter era to acceleration phase.}
\label{fig6}
\end{center}
\end{figure*}

\begin{table*}
\begin{center}
\caption{\bf The fixed points and physical parameters of the system (\ref{aoto}).}
\vspace{5mm}
\begin{tabular}{|c|c|c|c|c|}
  \hline

  {\bf fixed point}  & $\Omega_{m}$ & $\Omega_{r}$ & $X$&$q$      \\\hline\hline

 $P_{1}$ & $0$ & $0$ & $1$& $-1$   \\\hline

  $P_{2}$ & $0$ & $0$ & $\frac{3-\alpha}{6}$& $\frac{\alpha-1}{4}$    \\\hline

  $P_{3}$ & $0$ & $\frac{3\alpha(3-\alpha)}{2(10-3\alpha)}-\alpha+1$ & $\frac{\alpha(3-\alpha)}{2(10-3\alpha)}$& $1-\frac{\alpha}{2}$   \\\hline

  $P_{4}$ & $1-\alpha$ & $0$ & $\frac{\alpha}{3}$& $\frac{1-\alpha}{2}$  \\ \hline
\end{tabular}
\begin{tabular}{|c|c|c|c|c|c|}
\hline
{\bf fixed point}  &$\lambda_{1}$ & $\lambda_{2}$ & $\lambda_{3}$&{\bf stable regime} &{\bf unstable regime}\\\hline\hline
$P_{1}$ &$\alpha-3$ & $\alpha-4$ & $-\frac{\alpha +3}{2}$  &$-3<\alpha <3$&-----\\ \hline
$P_{2}$ &$\frac{3}{2}(\alpha-1)$ & $\frac{3\alpha-5}{2}$ & $\frac{\alpha+3}{2}$ &$\alpha <-3$&$ \alpha>\frac{5}{3}$\\ \hline
$P_{3}$ &$1$ & $\frac{5-3\alpha}{2}$ & $4-\alpha$  &-----&$\alpha<\frac{5}{3}$\\ \hline
$P_{4}$ &$-1$ & $3-\alpha$ & $\frac{3}{2}(1-\alpha)$ &$\alpha >3$ &-----\\ \hline
\end{tabular}

\label{tab3}
\end{center}
\end{table*}

\begin{table*}
\begin{center}
\caption{\bf Stable and unstable regimes of the parameter $\alpha$ for the system (\ref{aoto}).}
\vspace{5mm}
\begin{tabular}{|c|c|c|}
  \hline
\textbf{$\alpha$ regime}&\textbf{stable fixed point}&\textbf{unstable fixed point}\\\hline
 $\alpha<-3$&$P_2$&$P_3$\\\hline
  $-3<\alpha<\frac{5}{3}$&$P_1$&$P_3$\\\hline
   $\frac{5}{3}<\alpha<3$&$P_1$&$P_2$\\\hline
    $\alpha>3$&$P_4$&$P_2$\\\hline
\end{tabular}
\label{tab4}
\end{center}
\end{table*}

In last Section by solving Eq. (\ref{contin1}) for $\rho_m$ and $\rho_r$ and plugging them in the Friedman equations (\ref{ff1}) and (\ref{ff2}) a specific  form of the $g(T)$ function was obtained. Although we generalized our calculations in this section but there are still some restrictions on the form of the continuity equation and also the form of the $f(R,T)$ function.   In the next section we look at some aspects of symmetry in the $f(R,T)$ gravity  to obtain a more complete view of it.

\section{$f(R,T)$ and Noether symmetry}
One of the tools to find symmetries of the theory is Noether symmetry approach. By the Noether theorem we know that there is a conserved charge related to every continuous symmetry of the theory that could be used to find the cyclic variables and reduce the dynamics of the system. Let $\mathcal{L}$ be a lagrangian defined on tangent space $\mathcal{TQ}=\{q_{i},\dot{q_{i}}\}$. A vector field on the tangent space can be represented by:
\bea
X=\alpha^{i}(q)\frac{\partial}{\partial q^{i}}+\dot{\alpha^{i}}(q)\frac{\partial}{\partial \dot{q}^{i}}
\eea
where dot means derivative with respect to time and $\alpha^{i}$ are Noether functions. Lie derivative of lagrangian $\mathcal{L}$ in the direction of $X$ is defined as:
\bea
L_{X}\mathcal{L}=X\mathcal{L}=\alpha^{i}(q)\frac{\partial \mathcal{L}}{\partial q^{i}}+\dot{\alpha^{i}}(q)\frac{\partial \mathcal{L}}{\partial \dot{q}^{i}}
\eea
The condition:
\bea
L_{X}\mathcal{L}=0
\eea
implies that $\mathcal{L}$ is conserved along the direction of $X$ or $X$ is a symmetry of $\mathcal{L}$. We have Euler-Lagrange equations too:
\bea
\frac{d}{dt}\frac{\partial \mathcal{L}}{\partial \dot{q}^{i}}-\frac{\partial \mathcal{L}}{\partial q^{i}}=0
\eea
contracting the above equation with $\alpha^{i}$ we have:
\bea
\alpha^{i}\left(\frac{d}{dt}\frac{\partial \mathcal{L}}{\partial \dot{q}^{i}}-\frac{\partial \mathcal{L}}{\partial q^{i}}\right)=0
\Rightarrow \frac{d}{dt}\left(\alpha^{i}\frac{\partial \mathcal{L}}{\partial \dot{q}^{i}}\right)=L_{X}\mathcal{L}\quad
\eea
We see from above equation that if $X$ is a symmetry of $\mathcal{L}$ (i.e. $L_{X}\mathcal{L}=0$) the function:
\bea
\mathcal{A}=\alpha^{i}\frac{\partial \mathcal{L}}{\partial \dot{q^{i}}}\label{charg}
\eea
is the constant of the motion or conserved charge. This is the Noether theorem.
\\Now we want to find the consistent form of $f(R,T)$ by Noether symmetry and its conserved charge. General action of $f(R,T)$ theory is:
\bea
S=\int d^{4}x \sqrt{-g}\left\{\frac{1}{2k}f(R,T)+ \mathcal{L}_{m}\right\}\label{3}
\eea
 From above discussion about Noether symmetry it is obvious that we need point like lagrangian to impose Noether constraint on the theory. To this aim we use Lagrange multipliers to set $R$ and $T$ as constraints of the motion. Integrating (\ref{3}) we have:
\bea
S&=&2\pi^{2}\int dt \frac{a^{3}}{2k}\{f(R,T)-\lambda\left[R+6\left(\frac{\ddot{a}}{a}+\frac{\dot{a}^{2}}{a^{2}}+\frac{\kappa}{a^{2}}\right)\right]\nonumber\\
&-&\lambda'\left[T-(\rho-3p)\right]+2k\mathcal{L}_{m}\}
\eea
By varying the action with respect to $R$ and $T$, we have $\lambda=f_{R}$ and $\lambda'=f_{T}$ so:
\bea
S&=&2\pi^{2}\int dt \frac{a^{3}}{2k}\{f(R,T)-f_{R}\left[R+6\left(\frac{\ddot{a}}{a}+\frac{\dot{a}^{2}}{a^{2}}+\frac{\kappa}{a^{2}}\right)\right]\nonumber\\
&-&f_{T}\left[T-(\rho-3p)\right]+2k\mathcal{L}_{m}\}
\eea
Integrating by parts we have for point like lagrangian of $f(R,T)$ model:\\
\bea
& &2k\mathcal{L}=a^{3}\left\{f-f_{R}R-f_{T}\left[T-(\rho-3p)\right]\right\}\label{pointl}\\
&+&6a\dot{a}^{2}f_{R}+6a^{2}\dot{a}\dot{R}f_{RR}-6\kappa af_{R}+6a^{2}\dot{a}\dot{T}f_{RT}-2k\omega\rho a^{3}\nonumber
\eea
We want to find the consistent form of the function $f(R,T)$ by Noether symmetry for two cases;
\begin{enumerate}
  \item $f(R,T)=h(R)+g(T)$ and we assume conservation of energy momentum tensor so we have equation (\ref{abc}) for $g(T)$. we can find the functionality of $h(R)$ and time behavior of the scale factor $a(t)$ by Noether symmetry.
  \item $f(R,T)=R+g(T)$ but energy conservation does not exist. We find the functionality of $g(T)$ and time behavior of the scale factor $a(t)$ by Noether symmetry.
\end{enumerate}
\subsection{$f(R,T)=h(R)+g(T)$}
Plugging (\ref{ro}) into (\ref{pointl}) we have:
\bea
2k\mathcal{L}&=&a^{3}\left(f-f_{R}R-f_{T}T\right)-6\kappa af_{R}+6a^{2}\dot{a}\dot{R}f_{RR}\nonumber\\
&&+6a\dot{a}^{2}f_{R}-6\kappa af_{R}+6a^{2}\dot{a}\dot{T}f_{RT}\nonumber\\
&&+f_{T}\rho_{0}(1-3\omega)a^{-3\omega}-2k\omega\rho_{0}a^{-3\omega}\nonumber
\eea
The tangent space for the above lagrangian is $\mathcal{TQ}=\{a,\dot{a},R,\dot{R},T,\dot{T}\}$. As discussed above the generator of the Noether symmetry is:
\bea
X=\alpha\frac{\partial}{\partial a}+\beta\frac{\partial}{\partial R}+\gamma\frac{\partial}{\partial T}+\dot{\alpha}\frac{\partial}{\partial \dot{a}}+\dot{\beta}\frac{\partial}{\partial \dot{R}}+\dot{\gamma}\frac{\partial}{\partial \dot{T}}\quad\quad
\eea
The Noether symmetry exist if at list one of the functions $\alpha$, $\beta$ and $\gamma$ is different from zero. So we should solve the equation:
\bea
L_{X}\mathcal{L}=X\mathcal{L}=0
\eea
By replacing $\frac{d}{dt}$ in $\dot{\alpha}$, $\dot{\beta}$ and $\dot{\gamma}$ as bellow:
\bea
\frac{d}{dt}=\frac{d}{da}\dot{a}+\frac{d}{dR}\dot{R}+\frac{d}{dT}\dot{T}
\eea
and setting the coefficient of $\dot{a}^{2}$, $\dot{R}^{2}$, $\dot{a}\dot{R}$, etc. equal to zero we find the following equations:
\bea
&f_{RR}\partial_{R}\alpha=0& \label{n1}\\
&2a\alpha f_{RR}+a^{2}f_{RRR}\beta+\gamma a^{2}f_{RRT}+\partial_{a}\alpha a^{2}f_{RR}&\label{n2}\\
&+2\partial_{R}\alpha f_{R}a+\partial_{R}\beta a^{2}f_{RR}+a^{2}f_{RT}\partial_{R}\gamma=0&\nonumber\\
&\alpha f_{R}+\beta f_{RR}a+\gamma f_{RT}a+2f_{R}a\partial_{a}\alpha+a^{2}f_{RR}\partial_{a}\beta& \nonumber\\&+a^{2}f_{RT}\partial_{a}\gamma=0 &\label{n3}\\
&2\partial_{T}\alpha f_{R}a+a^{2}\partial_{T}\beta f_{RR}+2a\alpha f_{RT}+\beta a^{2}f_{RRT}&\label{n4}\\
&+\gamma a^{2}f_{RTT}+a^{2}f_{RT}\partial_{T}\gamma +a^{2}f_{RT}\partial_{a}\alpha=0&\nonumber\\
&\partial_{T}\alpha f_{RR}=0&\label{n5}
\eea
and Noether constraint:
\bea
&3\alpha a^{2}\left(f-f_{R}R-f_{T}T\right)-6\kappa f_{R}\alpha& \nonumber\\
&-3\alpha\omega\rho_{0}[f_{T}(1-3\omega)-2k\omega]a^{-3\omega-1}& \nonumber\\
&-\beta a^{3}\left(f_{RR}R+f_{TR}T\right)+\beta f_{RT}\rho_{0}(1-3\omega)a^{-3\omega}&\nonumber\\
& -6\beta \kappa f_{RR}a-\gamma a^{3}(f_{RT}R+f_{TT}T)&\nonumber \\
&-6\kappa\gamma f_{RT}a+\gamma f_{TT}\rho_{0}(1-3\omega)a^{-3\omega}=0& \label{nc}
\eea
from (\ref{n1}) and (\ref{n5}) we have:
\bea
\partial _{a}\alpha=\partial_{T}\alpha =0\Rightarrow \alpha=\alpha(a)\label{a}
\eea
by inserting the condition $f(R,T)=h(R)+g(T)$ and from (\ref{n4}) and (\ref{a}) we have:
\bea
\partial_{T}\beta =0\Rightarrow \beta =\beta (a,R)
\eea
We have two equations (\ref{n2}) and (\ref{n3}) for three functions $\alpha$, $\beta$ and $\gamma$. So one of the functions are free and we can choose any value for it. The most simple case is $\gamma =0$. So equations (\ref{n2}) and (\ref{n3}) become:
\bea
&2a\alpha f_{RR}+a^{2}f_{RRR}\beta+\frac{d\alpha}{da} a^{2}f_{RR}+\partial_{R}\alpha a^{2}f_{RR}=0&\label{n22}\quad\\
&\alpha f_{R}+\beta f_{RR}a+2f_{R}a\frac{d\alpha}{da}+a^{2}f_{RR}\partial_{a}\beta=0&\label{n33}
\eea
Same equations as in ref. \cite{fr1} is obtained. The solutions are:
\bea
\alpha =c_{1}a+\frac{c_{2}}{a}, \quad \beta =-\left(3c_{1}+\frac{c_{2}}{a^{2}}\right)\frac{f_{R}}{f_{RR}}+\frac{c_{3}}{af_{RR}}\quad\label{4}
\eea
where $c_{1},c_{2},c_{3}$ are constants. Plugging (\ref{4}) into (\ref{nc}) we have:
\bea
& (c_{1}a+\frac{c_{2}}{a})[3a^{2}\left(h-Rh'-\theta g_{0}T^{\theta}\right)&\nonumber\\
&-6\kappa h' -3\omega\rho_{0}\left[\theta g_{0}T^{\theta}(1-3\omega)-2k\omega\right]a^{-3\omega-1}]&\nonumber\\
& \left[-\left(3c_{1}+\frac{c_{2}}{a^{2}}\right)h'+\frac{c_3}{a}\right]\left(Ra^{3}-6\kappa a\right)=0&\label{n6}
\eea
from equation (\ref{charg}) the conserved charge of the theory is:
\bea
\mathcal{A}&=&6c_{1}a\left(h''a^{2}\dot{R}-h'\dot{a}\right)+6c_{2}(h''a^{2}\dot{R}+h'\dot{a})+6h'a\dot{a}c_3\nonumber
\eea
By substituting $\alpha$ and $\beta$ and the answer of equation (\ref{n6}) in the above equation we can find time behavior of the scale factor $a(t)$.
\subsection{$f(R,T)=R+g(T)$}
\begin{figure*}
\begin{center}
\begin{tabular}{cc}
  \includegraphics[scale=0.7]{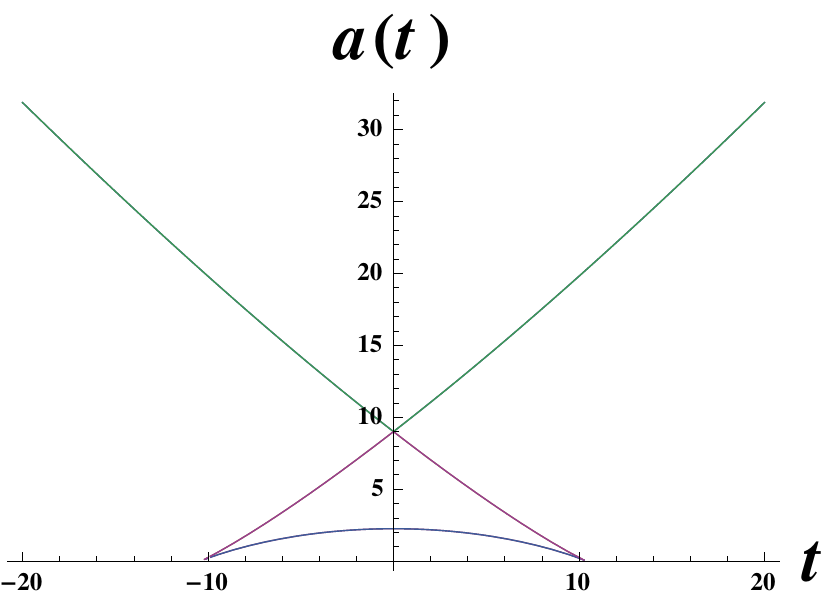}&\includegraphics[scale=0.7]{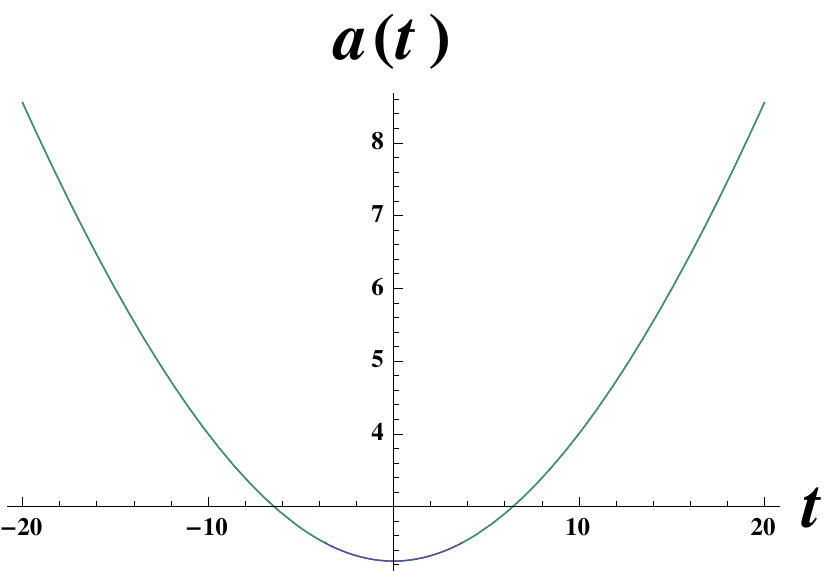}
\end{tabular}

\caption{ The behavior of $a(t)$ against $t$ by Noether symmetry in dS space (left figure) and AdS space (right figure)}
\label{fig7}
\end{center}
\end{figure*}

In this section we assume that energy momentum tensor is nonconserved. The point like lagrangian assuming $p=\omega\rho$ using equations (\ref{fr1}), (\ref{fr2}) and (\ref{pointl}) becomes:
\bea
2k\mathcal{L}&=&a^{3}(g-Tg')-6\kappa a+6a\dot{a}^{2}\nonumber\\
&+&\left[g'(1-3\omega)-2k\omega\right]\frac{3a\dot{a}^{2}-\frac{g}{2}a^{3}+\kappa a}{k+(1+\omega)g'}\label{n7}
\eea
where prime stands for differentiation with respect to the argument. $R$ is absent in the lagrangian (\ref{n7}) so the Noether symmetry is:
\bea
X\mathcal{L}=\alpha \frac{\partial \mathcal{L}}{\partial a}+\gamma \frac{\partial \mathcal{L}}{\partial T}+\dot{\alpha} \frac{\partial \mathcal{L}}{\partial\dot{a}}+\dot{\gamma} \frac{\partial \mathcal{L}}{\partial \dot{T}}=0
\eea
By imposing the Noether symmetry and setting the coefficients of $\dot{T}^{2}$, $\dot{a}^{2}$ and $\dot{T}\dot{a}$ equal to zero we find the following equations:\\
\bea
&\alpha +2a \partial_{a}\alpha =0&\label{n8}\\
&\partial_{T}\alpha =0&\label{n9}
\eea
that we have set $\gamma =0$ like previous section. From equations (\ref{n8}) and (\ref{n9}) $\alpha$ becomes:
\bea
\alpha =\frac{c}{\sqrt{a}}\label{alfa}
\eea
where $c$ is constant. Noether constraint and conserved charge are like bellow:
\bea
&(g-Tg')-\frac{2\kappa}{a^{2}}+\left(-\frac{1}{2}g+\frac{\kappa}{3\alpha a^{2}}\right)\frac{g'(1-3\omega)-2k\omega}{k+(1+\omega) g'}=0&\label{n10}\\
&\sqrt{a}\dot{a} \left[2+\frac{g'(1-3\omega)-2k\omega}{k+(1+\omega) g'}\right]=\mathcal{A}&\label{n11}
\eea
To see the consequence of the calculations presented in this section we try to solve a simple case of above equations. Assuming $\mathcal{A}=0$ in equation (\ref{n11}) $g(T)$ becomes:
\bea
g=g_{0}T\qquad g_{0}=\frac{-2k(1+\omega)}{5\omega +1}\label{n12}
\eea

It means that in high cosmological density limit ($\omega=1$) of the field equations by choosing $f(R,T)=R+g_0T$ we obtain conservation of energy and Noether symmetry simultaneously. As another consequence we can find scale factor; replacing (\ref{n12}) in (\ref{n10}) the following relation is found:
\bea
\kappa +\frac{\kappa}{3c}\sqrt{a}-\frac{1}{2}g_{0}(1-3\omega)\rho a^{2}=0
\eea
Replacing the term $\rho a^{2}$ from (\ref{fr1}) in above equation we have:
\bea
\kappa +\frac{\kappa}{3c}\sqrt{a}+\frac{(3\dot{a}^{2}+\kappa)(1-3\omega)(\omega+1)}{5\omega+1+(\omega-3)(\omega+1)}=0
\eea
The static universe is related to the flat case $\kappa =0$ or radiation ($\omega=1/3$). It means for radiation dominated or complete flat universe Noether symmetry leads to a static universe. Only a small curvature leads to a nonlinear differential equation above. In Fig. \ref{fig7} the behavior of $a(t)$ against $t$ is plotted for de Sitter (dS) and anti-de Sitter (AdS) spaces.

\section{Concluding remarks}
This paper dealt with the  $f(R,T)$ theory of gravity. We have studied equations of motion and future singularities for a barotropic perfect fluid and a dark energy like fluid assuming conservation of energy. To keep the conservation of stress-energy tensor, the choice of $f(R,T)$ is not completely arbitrary. It was found that there is no future singularity for the barotropic fluid while some kinds of singularity possibly exist for the dark energy like fluid  due to the new degrees of freedom in choosing the equation of state. We found relationships between the exponents of $t$ in the relations of $\rho$, $p$, $g$ and $g'$.  We showed that it is possible to explain the expansion of the universe by an effective running coupling constant where the pressure and density produced in the equations have the same behavior as the dark energy. \\
Considering singularities necessitates to study a special form of the scale factor. In order to generalize the studies we turned to the method of dynamical systems.  We found four fixed points which are related to radiation, matter and dark energy dominated accelerating phase of the universe. Behavior of the phase trajectories and evolution of the universe depends strongly on the value of $\alpha$ (the coupling constant of the interaction of the matter and radiation). There are four regimes in table \ref{tab4} that show different behaviors of the fixed points. In each regime evolution of the universe starts from un unstable point and ends in the stable one which is an accelerating fixed point in all cases. The preferred regime is $\frac{10}{3}<\alpha<3$ where matter and radiation fixed points are saddle points and $\Omega_r$ has positive values. It is interesting that weak energy condition is violated for $\Omega_m$ in this regime.\\
Finally, the effect of the Noether symmetry on $f(R,T)$ was studied and a consistent form of this function was determined using the Noether symmetry and the conserved charge for two cases. In the first one we assumed  $f(R,T)=f(R)+g(T)$  and also the  conservation of energy. In the second case we defined  $f(R,T)=R+g(T)$ with no requirement to conservation of energy momentum tensor. In both cases it is possible at list numerically to find the consistent form of the function $f(R,T)$  and time behavior of the scale factor $a$ simultaneously using the  Noether symmetry. In the he second case we can also have  both conservation of energy momentum tensor and Noether symmetry at the same time for $\omega=1$. For future research, it will be interesting to generalize this study to other types of gravitational theories.


\end{document}